\documentclass[prl,showpacs,amsmath,amssymb,twocolumn]{revtex4}

\usepackage{graphicx}
\newcommand{\Rth}{{\theta R_0}}
\newcommand{\Rla}{{\lambda R_0}}

\begin{document}

\title{Exterior complex scaling as a perfect absorber in time-dependent problems}

\author{Armin Scrinzi}

\email{armin.scrinzi@lmu.de}

\affiliation{Ludwig Maximilians University, Theresienstrasse 37, 80333 Munich, Germany, EU \\and\\ Wolfgang Pauli Institute, 1090 Vienna }


\date{\today}

\begin{abstract}
  It is shown that exterior complex scaling provides for complete
  absorption of outgoing flux in numerical solutions of the
  time-dependent Schr\"odinger equation with strong infrared
  fields. This is demonstrated by computing high harmonic spectra
  and wave-function overlaps with the
  exact solution for a one-dimensional model system and by
  three-dimensional calculations for the H atom and a Ne atom model.  
  We lay out the key ingredients for correct implementation and
  identify criteria for efficient discretization.
\end{abstract}

\maketitle


\section{Introduction}
The absorption of outgoing parts of the wave function at the boundaries 
of a finite volume is a key problem for any efficient numerical solution 
of the time-dependent Schr\"odinger equation  (TDSE) and it has been amply dealt with
also in recent literature (see, e.g., \cite{antoine08:absorption} and references therein). 
This interest has been renewed in the context
of intense laser-matter interactions: speaking in terms of physics, 
strong fields lead to large ionization and therefore large fluxes out of a 
central region. For strong field 
induced electronic and nuclear dynamics in atoms and molecules and high harmonic 
generation, electrons far from the system play 
no role and can be disregarded. 
When solving the TDSE for these processes, one can therefore 
identify an inner region (a finite volume) where an exact solution is
of interest. Outside that region one must, by some means, truncate the solution 
without compromising the inner region. This is particularly important for higher-dimensional
problems involving two or more electrons in order to control the size of the discretization. 
Out of the large number of approaches towards that goal the majority 
of computations of strong laser-matter interactions employed one of the following 
methods: absorbing masks \cite{krause92:mask}, complex absorbing potentials (CAPs) \cite{riss96:cap}, 
and exterior complex scaling (ECS)\cite{he07:complex_scaling,tao09:complex_scaling}.

The two recent numerical studies on ECS have cast doubt on the efficiency \cite{tao09:complex_scaling}  and maybe
even the fundamental correctness the method in numerical practice \cite{he07:complex_scaling}. 
In the present paper we will show that ECS indeed is a 
perfect absorber to full computational accuracy (14 digits). In addition, it allows highly 
efficient implementation where only a small fraction of the total discretization points are 
used for absorption. In both respects it far outperforms commonly used monomial CAPs.
As a third point, as noted early on \cite{mccurdy91:complex_scaling_tdse}, ECS is not just an absorber: ideally, it 
keeps a record of the dynamics in the outer region, which, in principle, could be recovered. 
We will provide numerical evidence for this fact. 

After giving a brief review of ECS, we will present with some care our discretization
method, as it plays an important role for correct and efficient implementation of ECS. 
The general characterization of ECS and a comparison with CAPs is done using a one-dimensional 
model system, and finally we will present results in three dimensions for the hydrogen atom 
and a single-electron model of Ne.

\section{TDSE with  a laser field}
We want to solve the TDSE of the general form
\begin{equation}
  \label{eq:tdse}
  i\frac{d}{d t} \Psi(\vec{x},t) = 
\left[
-\frac12 \Delta_{\vec{x}} + i\vec{A}(t)\cdot\vec{\nabla}_{\vec{x}} + V(\vec{x})
\right]\Psi(\vec{x},t),
\end{equation}
where $\vec{x}$ will be either a single $x$ or three $x,y,z$ spatial coordinates.
$\Delta_{\vec{x}},\vec{\nabla}_{\vec{x}}$ then denote $\frac{\partial^2}{\partial x^2},\frac{\partial}{\partial x}$
and Laplace and Nabla operators, respectively. $V(\vec{x})$ is a system-dependent binding 
potential and $\vec{A}(t)$ is the vector potential of the laser field. Here we have chosen the 
velocity gauge and removed the term $A(t)^2/2$ by a time-dependent unitary transform.
As the initial state we use the lowest energy eigenfunction of the field-free Hamiltonian 
operator $-\frac12 \Delta + V$. We use vector potentials with finite duration
\begin{equation}
  \label{eq:apot}
  \vec{A}(t)=\vec{A}_0 \cos^2\left(\frac{\pi t}{2nT}\right) \sin\left(\frac{2\pi t}{T}\right) 
\end{equation}
in the time interval $[-nT,nT]$ with $n=1,5,10$. The peak vector
potential is $\vec{A}_0=A_0$ and $\vec{A}_0=(A_0,0,0)$ in 1 and 3 dimensions, respectively. 
Such pulses with a single or a few oscillations of the electric field, linear polarization, 
and peak field amplitude at $t=0$ are frequently used as models in numerical studies.

The complete information of the system inside some inner region $|\vec{x}|\le R_0$ is
contained in the wave function amplitude. For characterizing the accuracy of our results 
by a single number, we use the overlap between an ``exact'' solution $\Psi_{\text{ex}}$ obtained 
from a calculation in a very large box and the approximate solution $\Psi$
\begin{equation}
  \label{eq:error_def}
  \mathcal{E}^2[B]=1-\frac{|\langle \Psi_{\text{ex}}|\Psi\rangle_B|^2}
{||\Psi_{\text{ex}}||^2_B ||\Psi||_B^2}.
\end{equation}
The scalar product is restricted to the inner region or a sub-region of the inner region 
$B\subset \left\{|\vec{x}|\le R_0\right\}$
\begin{equation}
  \label{eq:scalar}
  \langle \Psi_{\text{ex}}|\Psi\rangle_B=\int_B dx^{(d)} \Psi_{\text{ex}}^*(\vec{x})\Psi(\vec{x})
\end{equation}
and  $||\cdot||_B$ is the corresponding $\mathcal{L}^2$-norm.

A quantity of immediate physical interest is the intensity spectrum of the harmonic response 
given by the Fourier transform of the ``acceleration of the dipole''
\begin{equation}
  \label{eq:accel}
  \mathcal{S}_\Psi(\omega)=
\left\{
\mathcal{F}\left[\langle \Psi(\vec{x},t) | \frac{\partial V}{\partial x} |  \Psi(\vec{x},t) \rangle_B
\right]\right\}^2.
\end{equation}
For the comparison, integrals are restricted to the inner region $B$.
In general $\mathcal{S}(\omega)$ is a highly oscillatory quantity varying by several orders of magnitude. 
The local error of the spectrum relative to an ``exact'' spectrum is
\begin{equation}
  \label{eq:delta-harmon}
  \mathcal{D}(\omega) = 
\frac{\delta\omega[\mathcal{S}_\Psi(\omega) - \mathcal{S}_{\Psi_{\text{ex}}}(\omega)]}
{\int_{\omega-\delta\omega}^{\omega+\delta\omega}d\omega \mathcal{S}_\Psi(\omega)}.
\end{equation}
Local averaging in the denominator suppresses spurious spikes due to near-zeros of the spectrum.

\section{Outline of ECS theory}
\label{sec:ecs}
There is a large volume of literature available on complex scaling in general 
(see, e.g., 
\cite{balslev71:complex_scaling,aguilar71:complex_scaling,reed_simon82:complex_scaling}) and on exterior complex scaling 
in particular (see, e.g.,
 \cite{nicolaides78:complex_scaling,simon79:complex-scaling,mccurdy04:complex_scaling} and references therein). 
We restrict our summary to communicating the basic idea and to emphasizing the points that are 
essential for correct numerical implementation. For this we  closely follow earlier work, Ref.~\cite{scrinzi:jcp1993}.
In one dimension, exterior complex scaling consists in continuing the coordinate outside a ``scaling radius'' $R_0$
into the complex plane
\begin{equation}
  \label{eq:ecs-definition}
  x\to z_{\Rth}(x)\! =\! \left\{\begin{array}{lcr}
x&\text{for}&|x|<R_0\\
e^{i\theta}(x\pm R_0)\mp R_0&\text{for}& \mp x >R_0
\end{array}\right..
\end{equation}
The effect of the transformation on plane waves at values $x>R_0$ is
\begin{equation}
  \label{eq:ecs-planewave}
  e^{\pm ipx} \to e^{\pm ipR_0} e^{\pm ip \cos\theta (x-R_0)} e^{\mp p \sin\theta (x-R_0)}.
\end{equation}
For positive $p$ --- outgoing waves to the right side --- the functions become exponentially damped with 
increasing $x$, while for negative $p$ they grow exponentially. The corresponding situation with reversed
signs arises for $x<-R_0$. By complex scaling we can distinguish in- from outgoing waves
simply by their normalizibility without the need to analyze the asymptotic phase. 
In a typical discretization we implicitly or explictly use only square-integrable
functions, by which we exclude ingoing waves from a complex scaled calculation.
This is the key to complex scaling as a perfect absorber: in a well-defined region we have a simple 
instrument to systematically suppress ingoing waves by just requiring that our solution remain 
square-integrable. A more elaborate version of this reasoning can be found, e.g., in the appendix
of Ref.~\cite{scrinzi98}.

The mathematically rigorous theory of complex scaling often uses the alternative point of view that not
the wave functions, but rather the operator itself is scaled, while it remains an operator acting on 
an ordinary Hilbert space of square-integrable functions. We follow this line of reasoning for pointing 
out a fact of immediate computational relevance. We only give here plausibility arguments and refer the reader to 
Ref.~\cite{scrinzi:jcp1993} for a more extensive discussion and references to mathematical literature.

One starts from {\em real} scaling, i.e. replacing $i\theta$ in Eq.~(\ref{eq:ecs-definition}) with 
a real number $\lambda$ and observes that the transformation
\begin{equation}
  \label{eq:real-scaling}
  (U_{\Rla} \Psi)(x)\! =\! \left\{\begin{array}{lll}
\Psi(x)&\!\text{for}\!&x\!<\!R_0\\
e^{\lambda/2}\Psi(e^\lambda(x\!\mp\!R_0)\!\mp\! R_0)&\!\text{for}\!&\mp x\!>\!R_0\\
\end{array}
\right.
\end{equation}
is unitary. The scaling factor $e^{\lambda/2}$ is essential to ensure
unitarity. Formally, this transform can just as well be applied to the Hamiltonian by
defining
\begin{equation}
  \label{eq:real-scale-op}
  H_{\Rla}:=U_{\Rla}HU^*_{\Rla}.
\end{equation}
It is important to note that if $H$ is defined on differentiable functions $\Psi$, the transformed
operator is defined on the {\em discontinuous} functions $\Psi_\Rla=U_{\Rla}\Psi$
and its action on these functions is given by
\begin{equation}
  \label{eq:action}
  H_{\Rla}\Psi_\Rla = U_{\Rla}HU^*_{\Rla}\Psi_\Rla = U_{\Rla}H\Psi. 
\end{equation}
As a unitary transform $U_\Rla$ leaves the operator's spectrum unchanged and the
scaled dynamics is in a one-to-one relation to the unscaled.
Now the hard part of mathematical theory sets in: for a certain class of 
``dilation analytic'' potentials, 
the operators $H_{\Rla}$ can be analytically continued to complex values 
$\lambda\to i\theta$ without changes in the bound state spectrum \cite{reed_simon82:complex_scaling}. 
The continuous spectrum is rotated around the continuum threshold into the lower complex 
energy plane by the angle $2\theta$. This is trivial to see for the free particle 
and the case $R_0=0$ where the spectrum $\sigma(-\Delta)$ transforms as
\begin{equation}
  \label{eq:ecs-laplace-only}
  \sigma(-\Delta) = [0,\infty) \to \sigma(-e^{-2i\theta}\Delta) = [0,e^{-2i\theta}\infty).
\end{equation}
This property of the continuous spectrum persists when dilation analytic 
potentials are added and for $R_0>0$: the complex scaled Hamiltonian retains a 
distinct ``memory'' of the unscaled Hamiltonian. 
Proof of dilation analyticity can be difficult to find. 
Beyond some large $R_0$, where most physical potentials have simple decaying 
tails, the expected ECS properties are confirmed by numerical experiment.

One can now write an exterior complex scaled TDSE
\begin{eqnarray}
  \label{eq:ecs-tdse}
 \lefteqn{ i\frac{d}{d t} \Psi_{\Rth}(x,t) = H_\Rth(t) \Psi_\Rth(x,t)=}\nonumber\\
&& 
\left[
-\frac12 \Delta_{\Rth}\! +\! i\vec{A}(t)\cdot\vec{\nabla}_{\Rth}\! +\! V_\Rth(x)
\right]\!\Psi_{\Rth}(x,t).
\end{eqnarray}
Here it is assumed that the potential can be analytically continued to 
complex values $V_\Rth(x)=V[z_\Rth(x)]$.
One may hope that the dynamics generated by (\ref{eq:ecs-tdse}) is related to the original dynamics,
and that for $|x|<R_0$ the solution is identical to the unscaled solution
$\Psi_\Rth(\vec{x})=\Psi(\vec{x})$. We will demonstrate below that this 
expectation can be confirmed by numerical experiment to the level of full machine precision.

The main purpose of this brief discussion of ECS theory is to stress the importance of the
correct discontinuity in the wave function for defining differential operators.
The discontinuity at $R_0$ is intimately linked to the unitarity of the real scaled problem,
which in turn secures the conservation of spectral properties of the scaled operator. For given
$R_0$ and $\theta$ it has the explicit form
\begin{eqnarray}
  \label{eq:ecs-dicont}
  \Psi_{\Rth}(R_0-0)&=&e^{i\theta/2}\Psi_{\Rth}(R_0+0)\label{eq:ecs-discont1}\\
  \Psi'_{\Rth}(R_0-0)&=&e^{i3\theta/2}\Psi'_{\Rth}(R_0+0)\label{eq:ecs-discont2}.
\end{eqnarray}
The discontinuity condition (\ref{eq:ecs-discont2}) on the derivative  
arises from transforming the continuous first derivatives of the original functions. 
On such functions, one can define the complex scaled Laplacian in analogy to Eq.~(\ref{eq:action})
by ``back-scaling'' the scaled function $\Psi_{\Rth}(x)\to\Psi_{\Rth}(e^{-i\theta}(x\mp R_0)\pm R_0)$, applying the ordinary Laplacian, and forward-scaling the result:
\begin{equation}
  \label{eq:complex-scaling}
  (\Delta_{\Rth} \Psi_{\Rth})(x)\!=\! 
\left\{\begin{array}{lll}
\Delta\Psi_{\Rth}(x)&\text{for}&|x|\!<\!R_0\\
e^{-2i\theta}\Delta\Psi_{\Rth}(x)&\text{for}& |x|\!>\!R_0.\\
\end{array}
\right.
\end{equation}
The factor $e^{-2i\theta}$ appears,
as the derivative $\partial^2/\partial x^2$ is applied to the back-scaled function
rather than to $\Psi_{\Rth}(x)$.
On continuous functions the scaled Laplacian (and any derivative) is not defined as an operator 
in the Hilbert space, just as an ordinary Laplacian is not defined on discontinuous functions.

Finally we want to point to the fact that the adjoint operator 
$\left(\Delta_{[0,\theta]}\right)^\dagger=\Delta_{[0,-\theta]}$
requires functions with the complex conjugate condition of 
Eqs.~(\ref{eq:ecs-discont1},\ref{eq:ecs-discont2}).
This means that for our discretization by a basis set the discontinuities
{\em must not be conjugated} when going from ket- to bra-vectors.
We will show below how this can be easily implemented in a finite element basis. 

\section{Discretization}
For the discretization of ECS two points are important: (i) the correct implementation of the 
discontinuity and (ii) good approximation of analyticity. Both 
can be most conveniently accommodated in a finite element discretization of high rank.

We follow the implementation strategy laid out
in Ref.~\cite{scrinzi:jcp1993}: each coordinate axis is divided into $N$ elements
$[x_{n-1},x_n], n=1,\ldots,N$. On each element we 
choose a set of $p_n$ linearly independent 
functions that can be transformed to obey the conditions
\begin{equation}
  \label{eq:fem-conditions}
\begin{array}{l}
  f^{(n)}_1(x_{n-1})=f^{(n)}_{p_n}(x_n)=1\\
f^{(n)}_i(x_{n-1})=f^{(n)}_i(x_n)=0 \quad \text{else}
\end{array}
\end{equation}
We will call $p_n$ the ``rank'' of the finite element.
In principle any set of functions that obeys (\ref{eq:fem-conditions}) can be
used in a finite-element scheme. 
In practice we use real-valued polynomials which for enhancing
numerical stability we transform to 
\begin{equation}
  \label{eq:overlap}
  \int_{x_n-1}^{x_n}f^{(n)}_i(x)f^{(n)}_j(x) dx= m_i^{(n)}\delta_{ij}\quad\forall ij\not=1p_n,p_n1
\end{equation}
with normalization constants $m_i^{(n)}$.
For the element functions (\ref{eq:fem-conditions}) Dirichlet boundary 
conditions are implemented
by omitting the functions $f^{(1)}_1$ and $f^{(N)}_{p_N}$. 
Alternatively, on the leftmost and rightmost intervals we use
polynomials times an exponential $e^{\pm \alpha x}$ with $+$ and $-$ signs
on the intervals $(-\infty,x_1]$ and $[x_{N-1},\infty)$, respectively.
The conditions on the end element functions are 
\begin{equation}
  \label{eq:fem-conditions-infty}
\begin{array}{c}
f^{(1)}_i(x_1)=0\quad \text{except}\quad f^{(1)}_{p_1}(x_1)=1\\
f^{(N)}_i(x_{N-1})=0\quad   \text{except} \quad f^{(N)}_{1}(x_{N-1})=1.
\end{array}
\end{equation}
The exponent $\alpha$ can be adjusted for 
best performance, but its exact value was found to be uncritical for ECS.

The finite-element ansatz for the total wave function is as usual
\begin{equation}
  \label{eq:fem-ansatz}
  \Psi(x,t) = \sum_{n=1}^N \sum_{i=1}^{p_n} c^{(n)}_{i}(t) f^{(n)}_i(x). 
\end{equation}
By construction of the $f^{(n)}_i$, Eqs.~(\ref{eq:fem-conditions}), continuity across 
element boundaries is assured by demanding
\begin{equation}
  \label{eq:fem-continuity-c}
  c^{(n-1)}_{p_{n-1}}=c^{(n)}_1,\quad n=2,\ldots,N.
\end{equation}
Elementwise overlap and Hamiltonian matrices are 
\begin{eqnarray}
  \label{eq:fe-operator-s}
  S^{(n)}_{ij}&=&\int_{x_{n-1}}^{x_n} J(x) \left[f^{(n)}_i(x)\right]^* f^{(n)}_j(x) dx \\
  \label{eq:fe-operator-h}
  H^{(n)}_{ij}&=&\int_{x_{n-1}}^{x_n} J(x) \left[f^{(n)}_i(x)\right]^* H(t) f^{(n)}_j(x) dx, 
\end{eqnarray}
where $J(x)$ denotes the Jacobian function for integration over $x$.
The elementwise matrices are added into the overall discretized matrices $\hat{H}$ and $\hat{S}$
such that the last row and column of each elementwise matrix overlaps with the first row and column of the
following element (see Fig.~\ref{fig:h-assemble}), which is equivalent to setting the corresponding
coefficients equal, Eq.~(\ref{eq:fem-continuity-c}).
As always in finite element methods, continuity of the first derivative 
does not need to be imposed (see Ref.~\cite{scrinzi:jcp1993} for a more detailed discussion).
$\hat{H}$ and $\hat{S}$ are $M\times M$ matrices with 
\begin{equation}
  \label{eq:total-coefficients}
  M=\left\{
\begin{array}{l}
\sum_{n=1}^N p_n -N-1 \text{ for all } |x_n|<\infty \\  
\sum_{n=1}^N p_n -N+1 \text{ for } |x_0|=|x_N|=\infty. \\  
\end{array}\right.
\end{equation}

\begin{figure}
\begin{equation}
\nonumber
\left(
\begin{array}{ccccccc}
\ddots&\vdots&\vdots&\vdots&\vdots\\
\ddots&H^{(n-1)}_{p_{n-1}-1p_{n-1}}             &0    &0                             &0\\
\hdots& H^{(n-1)}_{p_{n-1}p_{n-1}}\!\!+\!H^{(n)}_{11} &\cdots&H^{(n)}_{_n}                   &0\\
     0&H^{(n)}_{21}                           &\ddots&H^{(n)}_{2p_n}                 &0\\
\vdots&\vdots                                &\ddots&\vdots                       &\vdots\\
     0&H^{(n)}_{p_n-11}                        &\cdots&H^{(n)}_{p_{n-1}p_n}           &0\\
     0&H^{(n)}_{p_n  1}                        &\cdots&H^{(n)}_{p_n p_n}\!\!+\!H^{(n+1)}_{11}&H^{(n+1)}_{12}\\
     0&0                                     &0     &H^{(n+1)}_{21}                 &\ddots      \\
\vdots&\vdots                                &\vdots&\vdots                        &\ddots\\
\end{array}
\right)
\end{equation}
\caption{\label{fig:h-assemble}
Placement of the elementwise block $H^{(n)}_{ij}$ in the overall Hamiltonian matrix $\hat{H}$ }
\end{figure}

For ECS we choose the scaling radii to coincide with the element boundaries $x_{n{\pm}}=\pm R_0$.
The scaled elementwise Hamiltonian matrices are evaluated by substituting 
in (\ref{eq:fe-operator-h})
the Jacobian $J(x)$ and the operator $H(t)$ with their ECS equivalents
\begin{equation}
  \label{eq:fe-operator}
H^{(n)}_{\Rth,ij}=\left\{
\begin{array}{lcl}
\int_{x_{n-1}}^{x_n} dx J f^{(n)}_i H f^{(n)}_j
&|x|< R_0\\
e^{i\theta}\int_{x_{n-1}}^{x_n} dx J_\Rth f^{(n)}_i
H_{\Rth} f^{(n)}_j
& |x|> R_0\\
\end{array}
\right.
\end{equation}
As we use {\em real} functions $f^{(n)}_i$ we could omit the
complex conjugation and the resulting matrices are
complex symmetric, {\it i.e} $H^{(n)}_{\Rth,ij}=H^{(n)}_{\Rth,ji}$.
The discontinuity (\ref{eq:ecs-discont1}) 
is brought into the system by the factor $e^{i\theta}$ for the integrals 
$|x|>R_0$: it amounts to multiplying all functions
$f^{(n)}_i$ outside the scaling radius 
by $e^{i\theta/2}$ and as the discontinuity does not get complex conjugated, 
the bra and ket discontinuity factors
do not cancel but multiply to $e^{i\theta}$. Like in the 
unscaled case, the continuity condition on the first derivative
(\ref{eq:ecs-discont2}) does not need to be imposed for finite elements.
The procedure for constructing the overall matrix $\hat{H}_{\Rth}$ is 
identical to the unscaled case.
Replacing $H(t)$ by $1$ results in the correct (non-hermitian) overlap matrix 
$\hat{S}_{\Rth}$ for the discretized problem. 
In practice, the matrices $\hat{H}_\Rth$ and $\hat{S}_\Rth$ are rarely set up explicitly, as applying the 
elementwise matrices to the corresponding sections of the coefficient vectors is far more efficient.

There are no specific issues for time propagating the discretized system
\begin{equation}
  \label{eq:tdse-discrete}
  \hat{S}_{\Rth}\frac{d}{dt}\vec{c} =   \hat{H}_{\Rth}(t) \vec{c}
\end{equation}
except maybe that very high accuracy was needed for our comparisons. 
If anything, ECS mitigates the well-known
stiffness problems for explicit time-integrators, as high kinetic energies are also associated
with large imaginary parts and decay rapidly. We use Runge-Kutta schemes with self-adaptive step size
and self-adaptive order up to order 7. Robust error control is achieved by single-to-double-step comparisons.
We obtain significant speedups of the propagation by removing states with 
very high eigenvalues of the field-free Hamiltonian from the simulation space by explicit projection.

\section{ECS for a 1-d problem}
We first investigate ECS for the 1-dimensional ``hydrogen atom'' with 
the model potential
\begin{equation}
  \label{eq:1d-potential}
  V(x) = -\frac{1}{\sqrt{2+x^2}},
\end{equation}
which gives the ground state energy $-1/2$. Here and below use
the peak vector potential $|A_0|=1.26$ and the optical period $T=104.8$. 
If interpreted as atomic units, these parameters correspond to peak intensity $2\times10^{14}W/cm^2$ and
wave length $760\,nm$. We will show results for FWHM of amplitude of $n=1$, 5 and 10 optical 
cycles and total pulse durations of 2,10, and 20 optical cycles, see Eq.~(\ref{eq:apot}). 
The classical quiver amplitude of an electron in this field is $A_0\times T/2\pi=21$ atomic units.
At the end of a single cycle pulse with this intensity around 20\% 
and after a 5 cycle pulse more than 80\%  of the electron probability 
have left the range [-40,40].

Within the framework of this model system we will answer the following questions: 
Can ECS be considered a perfect absorber? 
Can the scaling radius be put inside the range of the quiver motion, i.e. $R_0< 21$? 
Does ECS work for long pulses?
Which parameters determine the efficiency of ECS?  
How many discretization coefficients are needed?
How does ECS perform compared to monomial CAPs? 
Does ECS work for length gauge? 

\subsection{ECS is a perfect absorber}
We call an absorber perfect, if the error $\mathcal{E}[-R_0,R_0]$ defined
in Eq.~(\ref{eq:error_def}) is on the level of machine precision. 
As the ``exact'' result $\Psi_{\text{ex}}$ for comparison we use a unscaled 
calculation on a large box $[x_2,x_{N-1}]=[-1180,1180]$ with a total of $M=4801$ discretization coefficients
distributed over 120  elements with constant rank $p_n\equiv 41$. The elements are 
equidistant except for the infinite end elements $x_0=-\infty$ and $x_N=\infty$ 
with exponent $\alpha=0.5$, Eq.~(\ref{eq:fem-conditions-infty}). 

For ECS we use the parameters $\theta=0.5$ and $R_0=40$ and finite elements that
 up to $R_0$ are the same as in the unscaled calculation. In the scaled
ranges on either end of the axis we use infinite elements $(\infty,-R_0]$ and $[R_0,\infty)$ 
with $p_1=p_N=41$ and exponent $\alpha=0.5$. At this point, no attempt was made to 
minimize the number of coefficients used for absorption by optimizing the scaling parameters. 
Indeed, with 
the given parameters we obtain for the $\mathcal{L}^2$ errors at the end of the pulses $t=nT$
\begin{equation}
  \label{eq:errors}
  \mathcal{E}[-R_0,R_0] = 
\left\{\begin{array}{ll}
2\times 10^{-15}&\text{ for } n=1 \\ 
3\times 10^{-14}&\text{ for } n=5 \\ 
\end{array}\right.
\end{equation}

\begin{figure}[!ht]
  \centering \includegraphics[height=8cm,angle=-90]{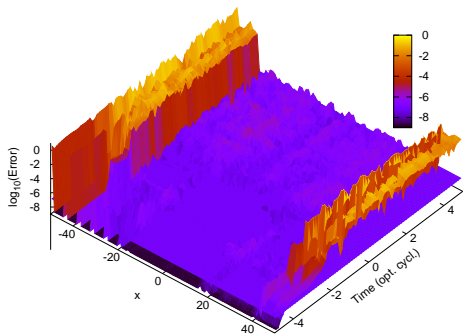}
  \caption{\label{fig:error}(color online)
Evolution of the relative error $|\Psi(x)-\Psi_{\text{ex}}(x)|/|\Psi(x)|$
during a 5-cycle pulse. 
The denominator is averaged over 5 grid points to avoid spurious spikes. 
For the pulse parameters and discretization, see text. 
The sharp rise of error marks the boundaries of the inner region.
A plane is drawn at error level $10^{-7}$ (color blue); only a few
error peaks in the inner region are above $10^{-6}$.
Away from the center, relative errors are enhanced initially as 
the wave function is nearly zero.}
\end{figure}

The error of the wave function amplitude is about the square root of these values and
it remains constant after the initial rise, see Fig.~(\ref{fig:error}).
The error level is constant through the whole range $[-R_0,R_0]$ and there
is a sharp edge to the scaled region, where the wave function is not directly 
related to the unscaled one. The errors indicate the accuracy limits of our numerical 
integration scheme and are not determined by ECS. It is therefore fair to say 
that, at least for the present model, ECS acts as a perfect absorber.

\subsection{Element rank and infinite end elements}
The choice of conspicuously high element rank for these very accurate 
calculations is not by coincidence. 
Complex scaling depends on analyticity properties of
the Hamiltonian. It is therefor not surprising that we observe a
strong dependency of the accuracy on the 
degree to which our discretization can approximate analytic functions.
Any localized basis, such as finite elements or B-splines 
is not analytic by definition because of the finite support of the basis functions. However
with increasing polynomial degree, loosely speaking, one gets closer
to analytic functions. Table~\ref{tab:degree} lists the error of the wave function
in the range [-35,35] for increasing element rank. 
As $\pm R_0$ must fall onto element boundaries, we had to choose slightly 
different values $R_0$ for the different element ranks. 
The ECS absorption range is discretized with between 
36 and 45 exponentially damped functions with $\alpha=0.4$ such that the sum of 
coefficients in the scaled and unscaled regions was $M=241$ for all calculations. 
For the error estimates at each $p_n$ a large box real calculations
was performed with same $p_n$ and the same number of points
as for ECS in $|x|<R_0$. 
From Table~\ref{tab:degree} we see that, depending on the desired accuracy, 
it is advisable to use polynomial degrees 8 or higher. 
\begin{table}[!ht]
\caption{\label{tab:degree}
Dependence of the final wave function error on the element rank $p_n$. 
All ECS calculations are for a single cycle pulse
and a total of $M=241$ discrete coefficients.}
\begin{tabular}{cccc} 
$p_n, n\not=1,N$ & $p_1$=$p_N$  & $R_0$ & $\mathcal{E}[-35,35]$ \\
 4&41 &40.  &  $4\times 10^{-8} $\\
 5&41 &40.  &  $1\times 10^{-7} $\\
 6&41 &40.  &  $3\times 10^{-10}$\\
 7&43 &39.  &  $1\times 10^{-9} $\\
 9&41 &40.  &  $5\times 10^{-12}$\\
11&41 &40.  &  $9\times 10^{-12}$\\
13&37 &42.  &  $2\times 10^{-13}$\\
15&46 &38.  &  $7\times 10^{-14}$\\
21&41 &40.  &  $2\times 10^{-15}$\\
\end{tabular}
\end{table}

For practical purposes we want to mention that the variation of an ECS
calculation with $\theta$ and box size is not a 
safe indicator of its accuracy. We found ECS calculations with fixed $R_0$, 
element rank and element sizes to be far better consistent among each 
other than their error relative to the unscaled result. For reliable
accuracy estimates one must vary the scaling radius $R_0$. 

The use of infinite elements at the ends of the simulation box  
greatly contributes to the good performance of ECS. 
Table~\ref{tab:finite} compares a few finite-box calculations with a calculation
using infinite end-elements with only 21 discretization points. Only at 
rather large finite boxes and a larger number discretization points 
one reaches the infinite element result. 
\begin{table}[!ht]
\caption{\label{tab:finite}
Error of ECS calculations with infinite and finite absorption
ranges. In all calculations we use a single-cycle pulse, 
rank $p_n=21$ and  160 discretization points in $[-R_0,R_0]=[-40,40]$. 
The length of the absorption range is $A=R_0-x_1=x_N-R_0$ and $M_A$ 
is the number of coefficients for absorption at each side.}
\begin{tabular}{cccc} 
$A$      & $M_A$ & $\mathcal{E}[-40,40]$ \\
$\infty$ & 21  &  $2\times 10^{-15} $\\
20      & 20  &  $4\times 10^{-4} $\\
40      & 40  &  $2\times 10^{-11}$\\
60      & 60  &  $1\times 10^{-9} $\\
80      & 80  &  $1\times 10^{-15}$\\
\end{tabular}
\end{table}

The explanation for this may be as follows: 
it was noticed in Ref.~\cite{tao09:complex_scaling} that long wave lengths cannot be 
accommodated in a finite ECS region and deteriorate absorption by reflections. 
Such long wave lengths have very little structure and should be 
easily parameterizable. It seems that the exponential 
tail of our end-element functions is sufficient to accommodate slowly varying
long wave-length parts of the ECS wave function. We leave a more 
detailed investigation of this observation to later work, but conclude that
efficient ECS is best done with infinite end intervals.

\subsection{Choice of  $R_0$ and back-scaling}
We find that the quality of the wavefunction in the unscaled region is 
not affected by the choice of the ECS radius $R_0$.  Table~\ref{tab:r0} shows
the errors $\mathcal{E}[-R_0,R_0]$ for $R_0=5,10,20$ and 40. The general
error level in these calculations is slightly higher as we used
a lower element rank of $p_n=11$ in order to be able to make the two elements of
the inner region small. The density of discretization points was kept constant
through all calculations. We see that the error level is independent
of whether the ECS radius is chosen inside $R_0=5,10,20$ or outside $R_0=40$
the classical quiver amplitude of $\alpha_0=21$.
Errors only start to rise, when the total size of the box indicated
by the number of discretization points $M$ becomes too small. 
This may not be surprising, if we assume that the spatial range of the dynamics remains
essentially unchanged by complex scaling: if the box, be it scaled or unscaled, cannot
let a particle go the full distance and then return without reflections, distortions
must occur. 

\begin{table}[!h]
\caption{\label{tab:r0} 
Dependence of the final wave function error on ECS radius $R_0$ and on the total number
of discretization points $M$.}
\begin{tabular}{cccc}
$M$ & $R_0$ &  $\mathcal{E}[-R_0,R_0]$ \\
241 & 40. &  $1.0\times 10^{-11} $\\ 
201 & 20. &  $5.6\times 10^{-12} $\\ 
160 & 10. &  $2.9\times 10^{-12}$\\ 
160 & 5.  &  $1.5\times 10^{-12} $\\
100 & 10. &  $1.8\times 10^{-12} $\\
80  & 10. &  $1.2\times 10^{-6}$\\ 
60  & 5.  &  $3.6\times 10^{-2}$\\  
\end{tabular}
\end{table}

There is an interesting conclusion that one may draw from this independence
of $R_0$: the fact that significant flux moves into the scaled region and 
then back out without corrupting the unscaled part of the
wave function indicates that also in the scaled region the TDSE
dynamics is encoded correctly, although in a different way.
Our numerical results are a striking corroboration of this conjecture 
that was made early on in ECS theory \cite{mccurdy91:complex_scaling_tdse}. In principle one may hope
to recover the unscaled wave function by analytic continuation. This hope
for back-scaling, in fact, was the original motivation for introducing the analytic form of 
functions on the end-elements, as an ordinary finite element function cannot
be unambiguously analytically continued. We have not further pursued this possibility
for two reasons: first, with larger $p_N=p_1$ and larger scaling angles $\theta$
we encountered severe numerical problems, as the back-scaled basis functions become
highly oscillatory and cancellation errors destroy the reconstruction of the unscaled
wave function; the second reason is the striking success of ECS with just a few
points needed for absorption. It is safer and simpler to just discard the small absorption
range and use the inner region directly for the evaluation of physical quantities.
Yet, if for one reason or another, one wishes to back-scale a time-dependent ECS wave 
function, our results indicate that such a procedure can be successful. One may in that
case use a representation of the scaled region that is less plagued by 
numerical problems than our exponential basis.

\subsection{Choice of scaling angle $\theta$ and exponent $\alpha$}
\begin{figure}[!ht] 
  \centering \includegraphics[height=8cm,angle=-90]{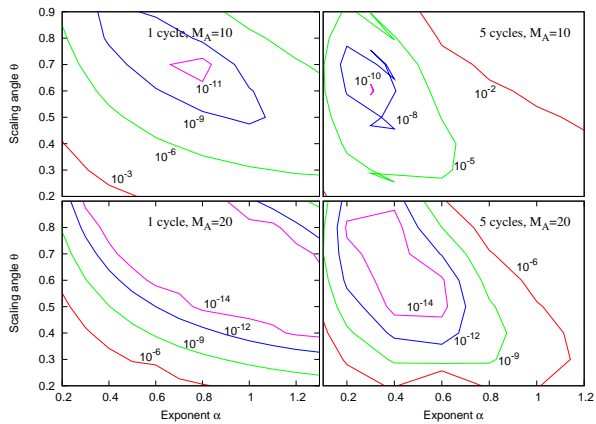}
  \caption{\label{fig:optimal}(color online)
Error $\mathcal{E}[-40,40]$ as a function of scaling angle $\theta$ and exponent $\alpha$ for n=1 and n=5-cycle pulses.
For the 5 cycle pulse, longer wave length reach the boundaries; optimal exponent and scaling angle are  
smaller and  a longer (20 point) absorption range significantly increases accuracy.}
\end{figure}

Although with sufficiently large absorption range one can always obtain perfect
absorption independent of scaling angle $\theta$ and damping exponent $\alpha$,
optimizing these parameters in a given situation allows to obtain good absorption 
with very few absorption points. Figure~\ref{fig:optimal} shows the error $\mathcal{E}[-40,40]$
for n=1 and n=5 cycle calculations with $M_A=10$ and 20 absorption points on either
end of the interval. The exact choice of the parameters is not critical for the $M_A=20$ calculations,
where full accuracy is reached for rather large parameter ranges.
As to be expected, the 5-cycle calculation with $M_A=10$ is most sensitive to $\theta$ and $\alpha$,
but still in a range of $\theta=\theta_0\pm 0.1$ and  $\alpha=\alpha_0\pm 0.1$ around the 
optimal values $\alpha_0,\theta_0\approx 0.3,0.6$  accuracy deteriorates only by 2 orders of 
magnitude to the still acceptable value of $10^{-8}$.
There is a clear anti-correlation 
between $\theta$ and $\alpha$, which may be explained looking at the oscillatory
behavior of the back-scaled exponential $\mathcal{I}m \exp\left[-\alpha r e^{-i\theta}\right]
=\sin\left[ \alpha \sin\theta r\right]$. 
We conjecture that the effective back-scaled wave number $\gamma=\alpha \sin\theta$ 
is the relevant parameter for efficient absorption. Correlation between the parameter $\gamma$ 
and $\theta$ nearly vanishes and optimization can safely be performed for each parameter independently.

\subsection{Comparison to complex absorbing potentials}
A popular and comparatively straight forward way of absorbing outgoing flux are 
complex absorbing potentials (CAPs). The basic idea is to add at the end of
the simulation box a potential with a negative imaginary part, which leads to 
exponential damping of the wave function. In this simplest form, the method
can be considered a differential form of absorption by mask functions, where
at preset intervals a certain part of the wave function is removed. 
A fundamental limitation of these methods is that they --- even in principle ---
cannot be strictly reflectionless. The attempt to obtain minimal 
reflections has lead to range of models, partially including real parts
into the potential and adjusting to specific physical situations (see, e.g. \cite{muga04:cap}).

It is beyond the scope of the present work to perform a comprehensive study 
of CAP for the present type of problems. 
Rather, we use the simple and 
well-investigated monomial CAPs \cite{riss96:cap}
\begin{equation}
  \label{eq:vcap}
  W(x)=-i\sigma x^q
\end{equation}
for polynomial degrees $q=4,6$ with optimized $\sigma$ in each calculation.
The criterion for our comparison with ECS is the number of discretization
points needed for a given level of absorption. 

Results are shown in Table~\ref{tab:cap}.
With a finite absorption range, ECS outperforms CAP roughly by one or two orders of magnitude.
However, when we use infinite end elements with ECS (discretized by only 21
points), we can reach absorption to machine precision. We could not find
a similar adjustment for CAP. 

\begin{table}[h]
\caption{\label{tab:cap} 
Accuracy of ECS and CAP for different absorption ranges $A$ and number of absorption coefficients $M_A$.
Scaling angle $\theta$ and absorption strength $\sigma$ 
for ECS and CAP, respectively, were optimized. The errors are calculated at the end of a single-cycle pulse.}
\begin{tabular}{cccccc}
Method&$M_A$ & $A$ & $\theta$ or $\sigma$ & q &$\mathcal{E}[-R_0,R_0]$ \\
ECS&21&$\infty$& 0.6  & --- & $2\times 10^{-15} $\\ 
ECS&20& 10     & 0.6  & --- &$2\times 10^{-4} $\\ 
ECS&40& 20 & 0.5 & --- & $1\times 10^{-7} $\\ 
CAP&20& 10 &        $10^{-4}$&4& $3\times 10^{-3} $\\ 
CAP&20& 10 & $2\times10^{-6}$&6& $4\times 10^{-3} $\\ 
CAP&40& 20 & $4\times10^{-6}$&4& $3\times 10^{-4} $\\ 
CAP&60& 30 & $6\times10^{-7}$&4& $1\times 10^{-5} $\\ 
\end{tabular}
\end{table}

\subsection{High harmonic spectra}
\begin{figure}[!ht] 
  \centering
 \includegraphics[height=8cm,angle=-90]{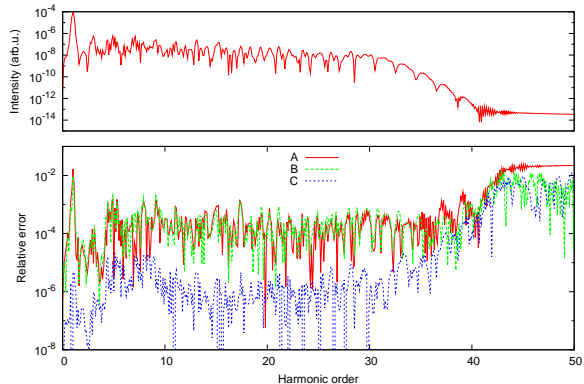}
  \caption{\label{fig:harmz}(color online) High harmonic power spectrum for a 5-cycle pulse (upper panel).
Lower panel: Accuracy $\mathcal{D}(\omega)$ 
of the high harmonic spectrum with different ECS parameters and discretizations.
Curve A is the error for $(R_0,M_A,\theta,\alpha)=(40,20,0.7,0.7)$ relative
to a fully converged real calculation. The choice of $R_0$ has has the largest influence
on accuracy: curve B, the difference between two calculations with $R_0=40$ 
and $R_0=50$ closely follows the overall error curve A. At fixed $R_0$ the influence
of the other ECS parameters and discretization is small: curve
C compares a calculations using  $(R_0,M_A,\theta,\alpha)$=(40,20,0.7,0.7) and finite element rank $p_n=21$ 
with (40,40,0.5,0.3) and rank 41.}
\end{figure}

Although the error $\mathcal{E}$ is a meaningful measure for wave function accuracy, it 
cannot be directly related to the error of a given observable. Figure~\ref{fig:harmz} shows
 the accuracy of ECS high harmonic spectra of 1- and 5-cycle pulses relative to a
real calculation. We find errors on the level between $10^{-4}$ and $10^{-3}$ and we could not
get much better agreement than this irrespective of discretization and  scaling parameters. Again this error is 
related to the numerical limits of our discretization and propagation schemes: the wave function error is 
$\sim 10^{-7}$ and the high frequency signal is suppressed by $10^{-4}$ relative to the 
fundamental peak, making a relative error of the suppressed signal of the order $10^{-3}$ quite
plausible. Indeed we find similar errors when comparing different, but equally accurate
purely real calculations. More disquieting is the $\sim 1\%$ error at the fundamental frequency,
which does not appear in large box real calculations.  We were not able to 
locate the origin of this error: it persists through variations of $R_0$, specific discretizations,
different time-discretizations, and also for the 3-d hydrogen calculation below (cf. Fig.~\ref{fig:hydrogen}). 
The error appears to be related to an artificial overall modulation
of the signal by the driving field, possibly related to internal normalizations during propagation.
Note that by construction normalization errors do not appear in the wave function accuracy measure $\mathcal{E}$,
Eq.~(\ref{eq:error_def}).
We believe, however, that this error is acceptable for all practical purposes.

\subsection{ECS fails in length gauge}
For field-interaction in length gauge
\begin{equation}
  \label{eq:length-gauge}
 i\vec{A}(t)\cdot\vec{\nabla}_{\vec{x}}  \to \vec{x}\cdot \frac{d\vec{A}}{dt}
\end{equation}
ECS completely fails in the time-dependent case. 
The reason for this behavior was pointed out in Ref.~\cite{mccurdy91:complex_scaling_tdse}:
when length gauge Volkov solutions are complex scaled their asymptotic behavior becomes dependent on
the sign of the field strength and alternates
between damping and growth. The convenient distinction between in- and outgoing
waves by their norms is lost. In the language of mathematical theory, $\vec{x}$ is not a dilation analytic
potential, and severely so: complex scaling transforms the spectrum of the Stark problem from purely continuous
into purely discrete and all discrete eigenvalues of the scaled Stark Hamiltonian have
imaginary parts \cite{herbst80:exponential_decay}. This is sharp contrast to dilation analytic potentials
where the bound state energies remain unchanged and the continuous spectrum 
is only rotated into the lower complex plane.

\section{Calculations for H and model  Ne}

In order to demonstrate the applicability of ECS to realistic problems, we show calculations
for the H atom with
\begin{equation}
  \label{eq:coulomb}
  V(\vec{x}) = -\frac{1}{|\vec{x}|}
\end{equation}
and a single electron model of the Ne atom with the potential
\begin{equation}
  \label{eq:neon}
  V(\vec{x}) = \sum_{i=1}^{4}a_i\frac{\exp[-c_i|\vec{x}|]}{|\vec{x}|}.
\end{equation}
We use the parameters given in Table~\ref{tab:neparam}, for which our model
reproduces the ground and first few excited state energies of Ne.
\begin{table}
\caption{\label{tab:neparam}Parameters for the Ne model potential Eq.~(\ref{eq:neon})}
\begin{tabular}{ccc}
& $a_i$ & $c_i$ \\
1&-1 &0\\
2&-0.3 &0.5\\
3&-2.05&2\\
4&1.23 &1\\
\end{tabular}
\end{table}
We use linearly polarized pulses with the same pulse shape and peak intensity as in the 
preceding section 
and pulse durations of 1 and 10 optical cycles. The calculations are done in polar coordinates
with a spherical harmonics basis on the angular coordinates and high rank finite elements
on the $r$-coordinate. Again an infinite last element is used.

\begin{figure}[!ht] 
  \centering
 \includegraphics[height=8cm,angle=-90]{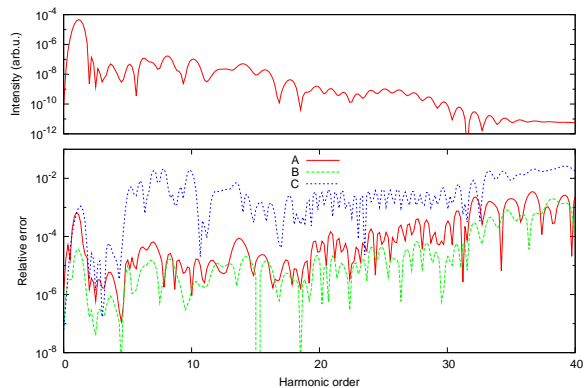}
   \caption{\label{fig:hydrogen} (color online) 
 High harmonic power spectrum from the hydrogen atom for a 1-cycle pulse (upper panel).
 Lower panel: Error  $\mathcal{D}(\omega)$  
with ECS parameters $(R_0,M_A,\theta,\alpha)=(40,20,0.5,0.5)$ relative
 to a $R_0=80$ calculation (curve A). The relative difference to a calculation with (40,40,0.4,0.4), curve B,
 underestimates the error. The calculation is converged with 20 angular momenta on the given level
of accuracy. More angular momenta do not change the result. At 15 angular momenta (curve C) accuracy
 deteriorates.}
\end{figure}

There are no surprises: convergence patterns and accuracy are very similar to the
one-dimensional model. Figure~\ref{fig:hydrogen} shows the harmonic
spectrum for hydrogen at 1 cycle together with errors for different ECS and discretization
parameters. Error estimate here is by comparison to an $R_0=80$ calculation.

\begin{figure}[!ht] 
  \centering
 \includegraphics[height=8cm,angle=-90]{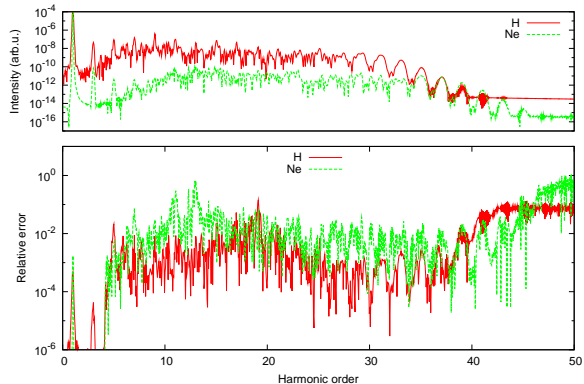}
  \caption{\label{fig:10cycle} (color online) 
High harmonic power spectra from the hydrogen and a neon model with a 10-cycle pulse (upper panel).
Relative accuracies shown in the lower panel are somewhat poorer with the longer pulse, 
in particular for Ne where due to the higher ionization potential the signal is very weak.}
\end{figure}

No new problems appear due to the more general Ne model potential (\ref{eq:neon}). Figure~\ref{fig:10cycle}
shows high harmonic spectra from a H and Ne for a 10-cycle pulse. Accuracy estimates
were obtained by varying ECS radius $R_0$.

\section{Discussion}
As we find high numerical stability and excellent performance of ECS as an absorber, the questions
arises what are the reasons for the numerical problems reported in 
Refs.~\cite{he07:complex_scaling,tao09:complex_scaling}, where
ECS was applied to essentially the same systems. One obvious source of inaccuracies may lie in
possibly low order discretizations. Unfortunately, in neither publication an investigation 
of the dependence of the observed effects on discretization is shown. 

Certainly the choice of finite box-sizes lowers the performance, but according to Table~\ref{tab:finite}
with the very large absorption ranges of 80 Bohr used in Ref.~\cite{tao09:complex_scaling}, excellent results should
be achievable. 

A possible source of the observed difficulties may be the treatment of the overlap matrix.
As discussed above we replace the ordinary overlap by the pseudo-overlap matrix $\hat{S}_\Rth$. 
With this choice and as we use strictly real finite element functions we obtain complex symmetric 
matrices
$(\hat{H}_\Rth)^T=\hat{H}_\Rth$ for zero field $A_0=0$ and $(\hat{S}_\Rth)^T=\hat{S}_\Rth$.
There are no explicit statements about $S_\Rth$ in 
Refs.~\cite{he07:complex_scaling,tao09:complex_scaling}.
Usually, finite difference methods imply (an approximation to) the identity operator for overlap.  
The B-spline method used in \cite{tao09:complex_scaling} requires a choice for $S_\Rth$ and 
Eq.~(20) of Ref.~\cite{tao09:complex_scaling} seems to imply that indeed the identity was used
as an overlap matrix. 

The comment on the non-orthogonality of the eigenvectors  of the non-normal
scaled Hamiltonian in  \cite{tao09:complex_scaling} also seems to indicate, 
that the ordinary, unscaled overlap matrix
$\hat{S}$ was used. Clearly, the eigenvectors $\vec{b}^{(\alpha)}$ of the eigenproblem
\begin{equation}
  \label{eq:eigen-ordinary}
  \hat{H}_\Rth \vec{b}^{(\alpha)} = \hat{S} \vec{b}^{(\alpha)} E_\alpha
\end{equation}
will not be orthogonal in general.
However, we find that all eigenvectors of the complex scaled generalized eigenproblem
\begin{equation}
  \label{eq:eigen-scaled}
  \hat{H}_\Rth \vec{c}^{(i)} = \hat{S}_\Rth \vec{c}^{(i)} E_i
\end{equation}
are pseudo-orthogonal and can be normalized in the sense
\begin{equation}
  \label{eq:pseudo-orthogonal}
  \sum_{lm} c^{(i)}_l \left(\hat{S}_\Rth\right)_{lm} c_m^{(j)}= \delta_{ij}.
\end{equation}
Then the matrix $\hat{H}_\Rth$ has a diagonal representation
\begin{equation}
  \label{eq:pseudo-spectral}
  \hat{H}_\Rth= \sum_i \vec{c}^{(i)} E_i \left(\vec{c}^{(i)}\right)^T
\end{equation}
and the spectral values $E_i$ appear as discrete approximations to the
true ECS spectrum with strictly non-positive imaginary parts $\mathcal{I}m E_\alpha\leq 0$. 
We do not have mathematical proof for this property of the discrete complex scaled system, 
but we find it valid in all our calculations on the level of computational accuracy. 
If on the other hand we use the ordinary overlap matrix $\hat{S}$,
we invariably obtain a few eigenvalues $E_\alpha$ with $\mathcal{I}m E_\alpha>0$
which will cause long-term instability of the time-propagation.
Possibly, this is the ultimate reason for the numerical instabilities observed in 
Refs.~\cite{he07:complex_scaling,tao09:complex_scaling}.

\section{Summary}
We have demonstrated that ECS can serve as a perfect absorber of outgoing flux in the 
sense that in the unscaled inner region it exactly matches a purely real 
calculation on a sufficiently large grid. 
We were able to push the agreement to relative $\mathcal{L}^2$ error of $10^{-15}$. 
The corresponding errors in the wave function amplitude are $\sim 10^{-7}$. 
Both errors are at the limits of our numerical integration
scheme. 

Furthermore we have evidence that also in numerical practice ECS does not 
just act as an absorber but conserves dynamical information during excursions into the
absorbing region: even when the quiver motion takes flux deeply into the ``absorbing''
region, the returning flux is identical to the flux in a purely real calculation.

For this, we found the following points essential: 
\begin{itemize}
\item[(i)] implementation of the correct scaled derivatives, including bra-functions with 
unconjugated discontinuity,
\item[(ii)] the use of ``infinite'' absorption ranges, which we discretized by 
polynomials times an exponential,
\item[(iii)]the use of high rank discretization also 
in the inner region to reach the highest accuracies.
\end{itemize}
Point (i) leads to a complex symmetric, in particular not positive definite discrete overlap 
matrix which must not be approximated by a positive definite matrix.

Following these rules, we encountered no numerical difficulties in the inner region or
in the absorbing region, using a standard explicit Runge-Kutta scheme for time integration. 
As a tendency, large scaling angles favor good absorption, in 
many cases we used $\theta=0.7\approx40^\circ$, which corresponds to an almost purely
imaginary continuous energy spectrum $[0,e^{-2i\theta}\infty)$. In our basis we found the scaling angle ultimately 
to be limited by numerical instabilities due to the complex symmetric overlap matrix.
As excellent absorption can be achieved with as few as 20 discretization coefficients in the absorbing
region, pushing the scaling angle to the numerical limits is not necessary in general
and scaling angles of $\theta=0.3\sim0.5$ served well for our purposes.
In general, we found the scheme numerically robust and not very sensitive to the 
scaling parameters. The option of back-scaling the solution to $\theta=0$ was abandoned
due to severe cancellation errors in the related transformations. 

When judging the accuracy of an ECS calculation, it is important to vary the 
ECS radius $R_0$. Our comparison with a real calculation indicates the variation
of the result with different $R_0$ gives realistic error estimates.
Other parameters such as rank of the finite elements, length of 
the absorption range, or scaling angle are of secondary importance.

ECS in the present implementation outperforms simple monomial CAPs. ECS errors 
were one or two order of magnitude smaller than CAP errors with the same spatial
discretization. When using infinite end intervals, the advantage of ECS can 
even reach 12 orders of magnitude!
We are aware of the fact CAPs can be greatly improved by a 
variety of measures (see, e.g., \cite{muga04:cap}). 
However, in general these require tuning of the CAP parameters
to a given situation. Even with that, we do not expect to reach comparable 
accuracies with CAPs as we could demonstrate for ECS.

We believe that ECS solves the absorption problem for the present class of systems
in any discretization, where implementation of (i)-(iii) is possible. 
Recovery of asymptotic information, such as electron spectra, may be inherently 
difficult as the total amount of information that is contained in the
scaled discretization is too small, which manifests itself in the ill-conditioning
of the back-scaling problem. However, in Ref.~\cite{caillat04:mctdhf} we presented a scheme 
for computing electron spectra under the assumption of a perfect absorber, at that point 
formulated for CAPs. An adaptation of that scheme to ECS and extension to few-body dynamics will
be investigated in future work.

\section{Acknowledgement}
This work was supported by the Viennese Science Foundation (WWTF) via the 
project "TDDFT" (MA-45).

\bibliography{/home/scrinzi/bibliography/photonics_theory}
\end{document}